\documentclass[11pt]{article} 
\usepackage[top=20mm,bottom=20mm, left=20mm, right=20mm]{geometry}

\usepackage{graphicx,amsmath,amssymb,url,enumerate,mathrsfs,subfig,epsfig,color,transparent,cite}
\usepackage{hyperref}
 \hypersetup{
     colorlinks=true,       
     linkcolor=red,          
     citecolor=blue,        
     filecolor=magenta,      
     urlcolor=cyan           
 }

\newcommand{\bp}[1]{\skew{5}\bar{#1}}

\title{Dislocations, disclinations, and metric anomalies as sources of global strain incompatibility in thin shells}
\author{Ayan Roychowdhury and Anurag Gupta\thanks{ag@iitk.ac.in}
\\Department of Mechanical Engineering, Indian Institute of Technology Kanpur, 208016, India.
}

\date{\today}

\begin{document}
\maketitle
\abstract{The strain incompatibility equations are discussed for nonlinear Kirchhoff-Love shells with sources of inhomogeneity arising due to a distribution of topological defects, such as dislocations and disclinations,  and metric anomalies, such as point defects, thermal strains, and biological growth. The incompatibility equations are given for all topological surfaces, with or without boundary, which are isometrically embeddable in a 3-dimensional Euclidean space.}

\section{INTRODUCTION}
For a sufficiently smooth strain field, defined over a fixed isometric embedding of the shell surface in a 3-dimensional Euclidean space $\mathcal{R}^3$, the global strain compatibility conditions ensure that there exist a global bijective deformation map, with respect to the fixed embedding, which is related to the strain field in a specified manner (as discussed below). This is indeed the case for a materially uniform, homogeneous, elastic shell, where the elastic strain, defined over a fixed stress-free reference configuration, is always compatible. A distribution of topological defects (dislocations and disclinations) and metric anomalies (point defects, heating, and growth) over the shell would however lead to material inhomogeneity and consequently to the non-existence of a connected stress-free configuration isometrically embeddable in $\mathcal{R}^3$. They would appear as sources of incompatibility in both local and global strain incompatibility equations \cite{rcgupta15,rcgupta17,zubov:book}.

The strain incompatibility equations are essential in order to calculate the internal stress field and the deformed shape of the shell, for a given distribution of defects and metric anomalies. They comprise of local relations in terms of partial differential equations, and global relations as integral equations, to be solved for incompatible strain fields. The global compatibility conditions include the integrability conditions for multiply connected surfaces such as cylinder, torus and M{\"o}bius band. In this article we collect all these relations for a nonlinear Kirchhoff-Love shell. The local relations have been recently derived for a more general nonlinear Cosserat shell model by \cite{rcgupta17}. The local relations, when reduced under the assumptions of von K{\'a}rm{\'a}n shell theory, generalize the incompatibility relations already existing in the literature for defective 2-dimensional crystals and fluid films \cite{bowickgiomi09}, 2-dimensional biological growth of shallow shells \cite{liang-mahadevan11}, and von K{\'a}rm{\'a}n plates \cite{zubov1}. The global relations, on the other hand, have appeared previously in the work of \cite[Chapter 5]{zubov:book}, see also \cite{zubov1}, but only for orientable surfaces. 
 
At first, the strain compatibility conditions, both local and global, are discussed in Section \ref{scc}. They are revisited to include defects and metric anomalies, and obtain the strain incompatibility relations, in Section \ref{inc}.

\section{STRAIN COMPATIBILITY CONDITIONS}
\label{scc}
Let $\omega$ be a connected, compact, 2-dimensional manifold, possibly with boundary, such that it is embeddable as a topological submanifold in $\mathcal{R}^3$. The manifold $\omega$ is our prototype for a thin shell. It is topologically characterized by its orientability, twistedness, the number of open discs removed, i.e., the boundaries, and other topological invariants; examples include open discs, spheres, with or without a finite number of handles, M{\"o}bius bands, among others. Let ${\bf R}(\theta^\alpha)$ be a global isometric embedding of $\omega$ into $\mathcal{R}^3$, such that $\omega$ is locally parametrized by coordinates $(\theta^1,\theta^2)$. With ${\bf A}_{\alpha} = {\bf R}_{,\alpha}$ and ${\bf N} ={{\bf A}_1\times{\bf A}_2}/{|{\bf A}_1\times{\bf A}_2|}$, we can construct the first and second fundamental forms associated with this embedding as $A_{\alpha\beta} = {\bf A}_{\alpha} \cdot {\bf A}_{\beta}$ and $B_{\alpha\beta} = - {\bf N}_{,\beta}\cdot{\bf A}_{\alpha}$, respectively. The Greek indices vary between 1 and 2, whereas the Roman indices vary between 1 and 3. The round and square brackets enclosing indices indicate symmetrization and anti-symmetrization, respectively, with respect to them.

\noindent \textit{The question of strain compatibility}: Given sufficiently smooth strain fields, $E_{\alpha\beta}$ (symmetric surface strain) and $\Lambda_{\alpha\beta}$ (transverse bending strain), over the fixed isometric embedding ${\bf R}(\theta^\alpha)$, what are the conditions to be satisfied for there to exist a sufficiently smooth global isometric embedding ${\bf r}(\theta^\alpha)$ of $\omega$ into ${R}^3$, with first and second fundamental forms $a_{\alpha\beta}$ and $b_{\alpha\beta}$ suitably constructed out of the given fields, such that 
\begin{equation}
E_{\alpha\beta} = \frac{1}{2}({\bf a}_{\alpha} \cdot {\bf a}_{\beta} -  {\bf A}_{\alpha} \cdot {\bf A}_{\beta}) =\frac{1}{2}(a_{\alpha\beta}-A_{\alpha\beta}) ~\text{and} \label{def3}
\end{equation}
\begin{equation}
\Lambda_{(\alpha\beta)} = {\bf n}_{,\beta} \cdot {\bf a}_{\alpha}- {\bf N}_{,\beta} \cdot {\bf A}_{\alpha} = -b_{\alpha\beta}+B_{\alpha\beta} \label{def4}
\end{equation}
everywhere on $\omega$, where ${\bf a}_{\alpha} ={\bf r}_{,\alpha}$ and ${\bf n} ={{\bf a}_1\times{\bf a}_2}/{|{\bf a}_1\times{\bf a}_2|}$? We answer this by first giving the local conditions followed by the global compatibility conditions. 

\noindent \textit{Local compatibility conditions}: From the given strain fields $E_{\alpha\beta}$ and $\Lambda_{\alpha\beta}$ we construct the two fundamental forms as $a_{\alpha\beta}  = A_{\alpha\beta}+2E_{\alpha\beta}$ and  $b_{\alpha\beta} = -\Lambda_{(\alpha\beta)}+B_{\alpha\beta}$. The local strain compatibility conditions, over a simply connected open set $W\subset \omega$, require $a_{\alpha\beta}$ to be positive-definite, $\Lambda_{[\alpha\beta]} =0$, 
  \begin{equation}
 \partial_1 b_{21}-\partial_2 b_{11} =0,~
 \partial_1 b_{22}-\partial_2 b_{12} =0,~\hbox{and} \label{compatibility6}
  \end{equation}
  \begin{equation}
 K_{1212}- (b_{12}^2  - b_{11}b_{22}) = 0,\label{compatibility7}
 \end{equation}
where $\partial$ denotes the covariant derivative with respect to the surface Christoffel symbols induced by the metric $a_{\alpha\beta}$ and $K_{1212}$ is the only independent component of the Riemann-Christoffel curvature associated with the Levi-Civita connection derived from $a_{\alpha\beta}$. The Gaussian curvature $K$ is such that $K_{1212} =  4aK$, where $a=\text{det} (a_{\alpha\beta})$. Equations \eqref{compatibility6} and \eqref{compatibility7} are the well-known Codazzi-Mainardi and Gauss equations for $a_{\alpha\beta}$ and $b_{\alpha\beta}$. These conditions are necessary and sufficient for the existence of a sufficiently smooth local isometric embedding ${\bf r}:W\to \mathcal{R}^3$, with first and second fundamental form given by $a_{\alpha\beta}$ and $b_{\alpha\beta}$, respectively, such that the PDEs in \eqref{def3} and \eqref{def4} are identically satisfied everywhere on $W$ \cite{ciar1}. 

\noindent \textit{Global compatibility conditions}: All 2-dimensional topological manifolds which are isometrically embeddable in $\mathcal{R}^3$ are homeomorphic to one of the following topological category of surfaces: (i) disc with no holes, (ii) disc with holes, (iii) 2-dimensional sphere $\mathcal{S}^2$, (iv) $\mathcal{S}^2$ with a finite number of attached handles, and (v) twisted bands. The last category, with odd number of twists, yields a non-orientable topological surface, e.g.,  M{\"o}bius band. Note that $\mathcal{S}^2$ is the only topology attainable by a compact, closed, simply connected 2-dimensional manifold embeddable in $\mathcal{R}^3$. We collect the global compatibility conditions for each of the categories below. While writing the conditions for cases (i), (ii), and (v), we will restrict our attention to the interior of the domain and therefore neglect any conditions that might be required at the boundary \cite{ciarletmardare2014}. 

\begin{figure}[t]
 \centering
 \includegraphics[scale=0.8]{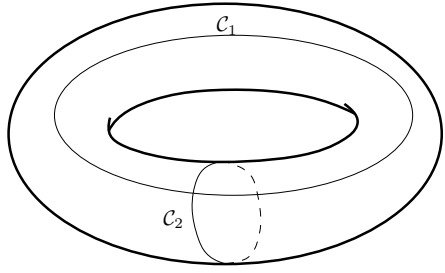}
 \caption{Irreducible loops on a toroidal shell.}
 \label{torus}
\end{figure}

The global integrability conditions, to be satisfied by strain fields, are given by 
\begin{equation}
  \prod (\mathcal{C}) e^{L|_{\mathcal{C}}(r)\,dr} ={\bf I}~\text{and}~
 \oint_{\mathcal{C}} {\bf a}_\alpha(s) d C^\alpha(s) ={\bf 0}
 \label{global}
\end{equation}
for every irreducible loop $\mathcal{C}$ in the domain. The former condition ensures the existence of a single-valued surface deformation gradient while the latter ensures the existence of a single-valued displacement map. Here, $L|_{\gamma}(s)=L_\alpha(s)\frac{d\gamma^\alpha(s)}{ds}$, where $\{\gamma^\alpha(s)\boldsymbol{A}_\alpha(s)\in\boldsymbol{R}(\omega),\,s\in[0,1]\}$ is an arc length parametrization of a curve $\gamma$ on the surface, and 
\begin{equation}
L_\alpha =\left[
 \begin{array}{ccc}
  \Gamma_{\alpha 1}^1 & \Gamma_{\alpha 1}^2 &  b_{\alpha 1}\\
  \Gamma_{\alpha 2}^1 & \Gamma_{\alpha 2}^2 &  b_{\alpha 2}\\
  - b_{\alpha}^1 & - b_{\alpha}^2 & 0
 \end{array}
\right],
\end{equation}
where $\Gamma^\tau_{\alpha\beta}=\frac{1}{2}a^{\tau\sigma}(a_{\sigma\alpha,\beta}+a_{\sigma\beta,\alpha}-a_{\alpha\beta,\sigma})$,  $[a^{\tau\sigma}]=[a_{\tau\sigma}]^{-1}$, and $b_\alpha^\beta = b_{\alpha\tau}a^{\tau\beta}$. Moreover, $\prod_0^s (\gamma) e^{L|_{\gamma}(s')\,ds'}={\bf I}+\int_0^s L|_{\gamma}(s')\,ds' + \int_0^s [\int_0^{s'} L|_{\gamma}(s'')\,ds'']ds'+\cdots$, where $\bf I$ is the 3-dimensional identity tensor. The vectors ${\bf a}_\alpha(s)$ in \eqref{global}$_2$ are obtained by solving $ [{\bf a}_1(s),{\bf a}_2(s),{\bf n}(s)]^T =  \prod_0^s (\gamma) e^{L|_{\gamma}(s')\,ds'}[{\bf a}_1(s_0),{\bf a}_2(s_0),{\bf n}(s_0)]^T$, where the values are known at some fixed point $s_0$. Equations \eqref{global}, when written for reducible loops, yield the local compatibility equations \eqref{compatibility6} and \eqref{compatibility7}. The number of irreducible loops is commensurate with the number of holes in the case of surfaces which are of category (ii). For a torus with genus one, which is the simplest example of category (iv), there are two mutually non-homotopic sets of irreducible loops, see Figure \ref{torus}. For twisted bands, the global condition \eqref{global}$_1$ is replaced by
\begin{equation}
  \prod (\mathcal{C}) e^{L|_{\mathcal{C}}(r)\,dr} ={\bf Q}^n,
 \label{globalt}
\end{equation}
where ${\bf Q}=\mbox{diag}[-1,1,-1]$ and $n$ is the number of twists. There are no non-trivial global integrability conditions for the simply connected surfaces of category (i) and (iii). The global conditions \eqref{global} combined with the local compatibility ensures that a smooth diffeomorphism exists between the current and the reference surface of the shell. This will also require the deformed shell to naturally imbibe the same topology as that of the reference shell. The latter is assumed to be known and hence there are no additional topological restrictions on the fundamental forms $a_{\alpha\beta}$ and $b_{\alpha\beta}$ or equivalently on strains. The global conditions \eqref{global} have previously appeared in \cite[Chapter 5]{zubov:book} but only for orientable surfaces, see also \cite{PietraszkiewiczVallee07}. 

\section{STRAIN INCOMPATIBILITY DUE TO A DISTRIBUTION OF DEFECTS}
\label{inc}

The strain fields which do not satisfy all the compatibility conditions are called incompatible.  Incompatible strains are well known sources of residual stress and deformation in continuum structures. The fields $a_{\alpha\beta}$ and $b_{\alpha\beta}$, constructed out of the incompatible strain fields, do not correspond to the first and second fundamental form of any realizable isometric embedding of $\omega$ into ${R}^3$, not even locally.  In this section, our aim is to characterize strain incompatibility in terms of defect distributions and metric anomalies. 

\begin{figure}[t]
 \centering
 \includegraphics[scale=0.7]{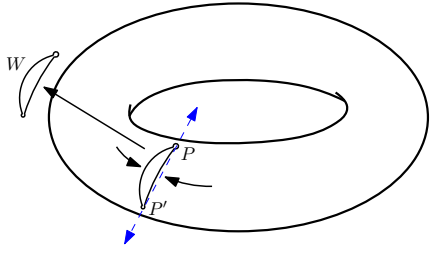}
 \caption{A local wedge disclination at a point $P$ on a torus and an associated antipodal disclination at $P'$.}
 \label{volterra-torus}
 \end{figure}

\begin{figure}[t]
\centering
\includegraphics[scale=0.7]{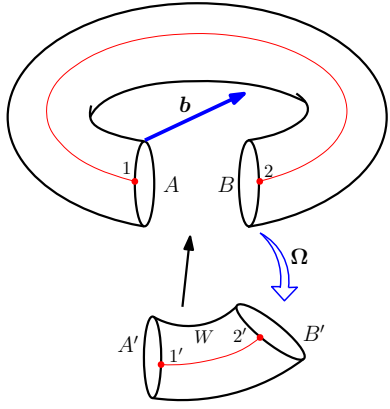}
\caption{Insertion of a wedge with non-parallel faces into a torus produces a global anomaly consisting of a global dislocation and a global wedge disclination.}
\label{global-defects-torus}
\end{figure}

\noindent \textit{Defect densities}: We assume the presence of in-surface dislocations densities, denoted by $J^\alpha$ for edge and $J^3$ for screw dislocations, in-surface wedge disclinations $\Theta^3$,  and metric anomaly density $Q_{ijk}$. Hence we ignore the out-of-surface dislocations and disclinations and the in-surface twist disclinations. Whereas the in-surface defects are commonly observed in 2-dimensional crystalline structures, the out-of-surface defects may be present in layered shells. More information about the geometric nature of these densities and the restrictions imposed on them by the Bianchi-Padova relations can be seen from \cite{rcgupta17}. We note, in particular, that the metric anomaly density can always be written in terms of a strain like second-order tensor (representing thermal or growth strains) whenever disclinations do not occur  \cite{rcgupta16}. It should also be noted that the defect lines, in case of local defects, always pass through the surface, see Figure \ref{volterra-torus}. This is unlike the global defects, where the defect lines always fall out of the surface, e.g. Figure \ref{global-defects-torus}, \cite[Chapter 1]{thesis-ayan}. It is also clear from Figure \ref{volterra-torus} that isolated defects on closed and compact surfaces will always have an associated antipodal counterpart.

\noindent \textit{Local incompatibility relations}: The local strain incompatibility relations for Kirchhoff-Love shells under these considerations are given by \cite{rcgupta17}
{\small \begin{equation}
  K_{1212}+(b_{11}b_{22}-b^2_{12}) = a\Theta^3 -2\partial_{[1}W_{2]12} -2W_{[1|i 2|}\, W_{2]1}{}^i \label{strain-incompatibility-relations1-kl},
  \end{equation}
  \begin{equation}
-\partial_2 b_{1 1}+\partial_1 b_{1 2}= -2\partial_{[1}W_{2]1 3} -2W_{[1|i 3|}\, W_{2]1}{}^i, \label{strain-incompatibility-relations2-kl}
\end{equation}
\begin{equation}
-\partial_2 b_{2 1}+\partial_1 b_{2 2}=  -2\partial_{[1}W_{2]2 3} -2W_{[1|i 3|}\, W_{2]2}{}^i,~\text{and}\label{strain-incompatibility-relations3-kl}
\end{equation}
\begin{equation}
a^{\alpha\beta}\varepsilon_{\alpha\rho}\varepsilon_{\beta\sigma} a^{-1} (\Lambda^p_{[12]})^2 = -\partial_{\rho} M_{3\sigma}{}^3 - (C_{\rho\alpha}{}^3+M_{\rho\alpha}{}^3)\times\nonumber
\end{equation}
\begin{equation}
~~~~~~ (C_{3\sigma}{}^\alpha+M_{3\sigma}{}^\alpha) -M_{\rho 3}{}^3 M_{\sigma 3}^3 + M_{3\alpha}{}^3 W_{\rho\sigma}^\alpha + M_{33}{}^3 W_{\rho\sigma}^3, \label{strain-incompatibility-relations4-kl}
\end{equation}}%
where $W_{ij}{}^k$ are components of a tensor defined as a sum of the contortion and non-metricity tensors, $W_{ij}{}^k=C_{ij}{}^k+M_{ij}{}^k$. The components of the contortion tensor take a simple form $C_{3\beta}{}^3 = C_{\beta 3}{}^3= C_{3 3}{}^i = C_{3\beta 3} = C_{\beta 3 3}= C_{3 3 i} = 0,~C_{3\beta}{}^\alpha = C_{\beta 3}{}^\alpha = a^{\alpha\nu}\varepsilon_{\nu\beta}J^3$, $C_{3\beta\alpha} = C_{\beta 3\alpha} = C_{\alpha\beta}{}^3 =C_{\alpha\beta 3} = \varepsilon_{\alpha\beta}J^3$,
 $C_{\alpha\beta\mu} = J^\sigma\big( a_{\sigma\beta} \varepsilon_{\mu\alpha}  + a_{\sigma\alpha} \varepsilon_{\mu\beta} + a_{\sigma\mu} \varepsilon_{\alpha\beta}\big)$, and $C_{\alpha\beta}{}^\mu = a^{\mu\nu} C_{\alpha\beta\nu}$.
Here, $\varepsilon_{\alpha\beta}=a^{\frac{1}{2}}e_{\alpha\beta}$ and $e_{\alpha\beta}$ is the 2-dimensional permutation symbol. On the other hand, the components of the tensor associated with non-metricity are $M_{33}{}^3 = M_{333} = \frac{1}{2}Q_{333},~M_{33\alpha} = \frac{1}{2}(2Q_{3\alpha 3}-Q_{\alpha 33}),~M_{33}{}^\alpha = a^{\alpha\beta} M_{33\beta}$, $M_{3\alpha}{}^3 = M_{\alpha 3}{}^3 =M_{3\alpha 3} = M_{\alpha 33} =  \frac{1}{2} Q_{\alpha 33}$, $M_{3\alpha}{}^{\beta} = M_{\alpha 3}{}^\beta = \frac{1}{2} a^{\beta\nu} (Q_{3\nu\alpha}-Q_{\nu\alpha 3} + Q_{\alpha 3 \nu})$, $M_{\alpha\beta}{}^3 = M_{\alpha\beta 3} = \frac{1}{2}(Q_{\alpha 3\beta}-Q_{3\beta\alpha} + Q_{\beta\alpha 3})$, $M_{\alpha\beta\mu} =  \frac{1}{2}(Q_{\alpha\mu\beta}-Q_{\mu\beta\alpha}+ Q_{\beta\alpha\mu})$, and $M_{\alpha\beta}{}^\mu = a^{\mu\nu}M_{\alpha\beta\nu}$. 
 
We can reduce the above system of equations for the von K{\'a}rm{\'a}n shell theory by restricting to small surface strain but moderate rotation such that $E_{\alpha\beta}=O(\epsilon)$ and $\Lambda_{\alpha\beta}=O(\epsilon^{\frac{1}{2}})$, where $\epsilon$ is of the order of shell thickness \cite{NaghdiVongsarnpigoon1983}. Additionally, we assume that the density of wedge disclinations $\Theta^3$, and the density of edge dislocations $J^\alpha$, up to their first spatial derivatives, are of order $O(\epsilon)$. We take $J^3=0$ and retain only in-surface metric anomalies, whose densities are represented by the functions $Q_{\alpha
\mu\nu}$. These functions, along with their first spatial derivatives, are also assumed to be of order $O(\epsilon)$. These order assumptions are motivated from the assumed order of in-surface strain.
Thus the only non-trivial incompatibility equation is given by \cite{rcgupta17}
{\small \begin{equation}
  \bp\partial_{11}E_{22}+\bp\partial_{22}E_{11}-2\bp\partial_2\bp\partial_1 E_{12} +\Lambda^p_{11}\Lambda^p_{22}-(\Lambda^p_{12})^2 \nonumber
  \end{equation}
  \begin{equation}
  = A\Theta^3 + 2\sqrt{A}\, A_{\sigma[1} \bp\partial_{2]}J^\sigma -\bp\partial_{1} M_{212}+\bp\partial_{2} M_{112},\label{strain-incompatibility-relations1-kl-1-dislo}
\end{equation}}
where $\bp\partial$ denotes covariant derivative with respect to the surface Christoffel symbols induced by the metric $A_{\alpha\beta}$. This equation generalizes the form of incompatibility equation that has been used previously in the mechanics of 2-dimensional defective crystals \cite{bowickgiomi09}, blooming of a flower \cite{liang-mahadevan11}, and mechanics of defective linear shell \cite{zubov2} and von K{\'a}rm{\'a}n plate \cite{zubov1}. 

\begin{figure}[t]
 \centering
 \includegraphics[scale=0.8]{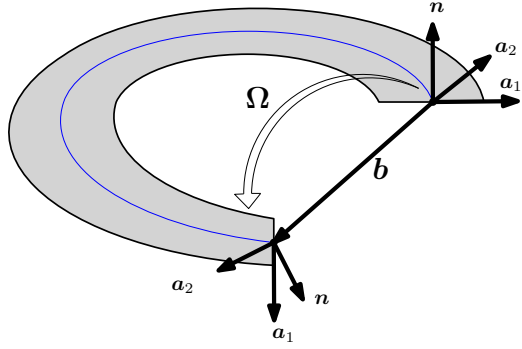}
 \caption{Disc with a hole after making a cut. The global Frank tensor $\boldsymbol{\Omega}$ and the global Burgers vector ${\bf b}$ measures the global incompatibility of the surface.}
 \label{disc-with-hole-cut}
\end{figure}

\noindent \textit{Global incompatibility relations}: The global incompatibility relations, assuming that the strains are locally compatible, are given by
\begin{equation}
  \prod (\mathcal{C}) e^{L|_{\mathcal{C}}(r)\,dr} =\boldsymbol{\Omega}({\mathcal{C}})~\text{and}
  \label{cond-global-incompatibility:1}
\end{equation}
\begin{equation}
 \oint_{\mathcal{C}} {\bf a}_\alpha(s) d C^\alpha(s) ={\bf b}(\mathcal{C}).
 \label{cond-global-incompatibility:2}
\end{equation}
 In the case of twisted bands, \eqref{cond-global-incompatibility:1} should be replaced by 
\begin{equation}
  \prod (\mathcal{C}) e^{L|_{\mathcal{C}}(r)\,dr} =\boldsymbol{\Omega}({\mathcal{C}}) {\bf Q}^n.
  \label{cond-global-incompatibility:3}
\end{equation}
Here, $\boldsymbol{\Omega}({\mathcal{C}})$, a second-order orthogonal tensor, is the global Frank tensor associated with the irreducible loop $\mathcal{C}$, see Figures \ref{global-defects-torus} and \ref{disc-with-hole-cut}. The orthogonality of $\boldsymbol{\Omega}({\mathcal{C}})$ is due to the fact that there are no distributed metrical disclinations on the surface, hence, inner product is preserved during the parallel transport \cite{rcgupta16}.
On the other hand, ${\bf b}(\mathcal{C})$ is the global Burgers vector associated with the loop $\mathcal{C}$, see Figures \ref{global-defects-torus} and \ref{disc-with-hole-cut}. These global conditions have been discussed by \cite[Chapter 5]{zubov:book} for surfaces homeomorphic to a disk with holes. A derivation of the global incompatibility conditions, with non-zero local incompatibility, remains an open problem. 

\bibliographystyle{plain}

\bibliography{ssta}

\begin{thebibliography}{10}

\bibitem{bowickgiomi09}
M~J Bowick and L~Giomi.
\newblock Two-dimensional matter: order, curvature and defects.
\newblock {\em Advances in Physics}, 58:449--563, 2009.

\bibitem{ciar1}
P~G Ciarlet.
\newblock An introduction to differential geometry with applications to
  elasticity.
\newblock {\em Journal of Elasticity}, 78-79:1--215, 2005.

\bibitem{ciarletmardare2014}
P~G Ciarlet and C~Mardare.
\newblock Intrinsic formulation of the displacement-traction problem in
  linearized elasticity.
\newblock {\em Mathematical Models and Methods in Applied Sciences},
  24:1197--1216, 2014.

\bibitem{liang-mahadevan11}
H~Liang and L~Mahadevan.
\newblock Growth, geometry, and mechanics of a blooming lily.
\newblock {\em Proceedings of the National Academy of Sciences in the United
  States of America}, 108:5516--5521, 2011.

\bibitem{NaghdiVongsarnpigoon1983}
P~M Naghdi and L~Vongsarnpigoon.
\newblock A theory of shells with small strain accompanied by moderate
  rotation.
\newblock {\em Archive for Rational Mechanics and Analysis}, 83:245--283, 1983.

\bibitem{PietraszkiewiczVallee07}
W~Pietraszkiewicz and C~Vall{\'e}e.
\newblock A method of shell theory in determination of the surface from
  components of its two fundamental forms.
\newblock {\em Zeitschrift f{\"u}r Angewandte Mathematik und Mechanik},
  87:603--615, 2007.

\bibitem{thesis-ayan}
A~Roychowdhury.
\newblock {\em Geometry and Mechanics of Defects in Structured surfaces}.
\newblock PhD thesis, Indian Institute of Technology Kanpur, 2017.

\bibitem{rcgupta15}
A~Roychowdhury and A~Gupta.
\newblock Material homogeneity and strain compatibility in thin elastic shells.
\newblock {\em Mathematics and Mechanics of Solids}, 10.1177/1081286515599438,
  2015.

\bibitem{rcgupta16}
A~Roychowdhury and A~Gupta.
\newblock Non-metric connection and metric anomalies in materially uniform
  elastic solids.
\newblock {\em Journal of Elasticity}, 126:1--26, 2017.

\bibitem{rcgupta17}
A~Roychowdhury and A~Gupta.
\newblock On structured surfaces with defects: geometry, strain
  incompatibility, stress field, and natural shapes.
\newblock {\em Under review, arxiv.org/abs/1702.03737}, 2017.

\bibitem{zubov:book}
L~M Zubov.
\newblock {\em Nonlinear Theory of Dislocations and Disclinations in Elastic
  Bodies}.
\newblock Springer, 1997.

\bibitem{zubov1}
L~M Zubov.
\newblock Von {K}\'arm\'an equations for an elastic plate with dislocations and
  disclinations.
\newblock {\em Doklady Physics}, 52:67--70, 2007.

\bibitem{zubov2}
L~M Zubov.
\newblock The linear theory of dislocations and disclinations in elastic
  shells.
\newblock {\em Journal of Applied Mathematics and Mechanics}, 74:663--672,
  2010.

\end{thebibliography}

\end{document}